\documentclass[journal]{IEEEtran}

\usepackage{xcolor,soul,framed} 

\colorlet{shadecolor}{yellow}
\usepackage[pdftex]{graphicx}
\graphicspath{{../pdf/}{../jpeg/}}
\DeclareGraphicsExtensions{.pdf,.jpeg,.png}

\usepackage[cmex10]{amsmath}
\usepackage{array}
\usepackage{mdwmath}
\usepackage{mdwtab}
\usepackage{eqparbox}
\usepackage{url}
 \usepackage{mathrsfs}
 \usepackage[bookmarks=false]{hyperref}
 \usepackage{xcolor}
 \usepackage{booktabs}
 \usepackage{array}
 \usepackage{ulem} 
\usepackage{cleveref}
\usepackage{cite}
\usepackage{multirow}
\usepackage{cleveref}

\begin{document}
\bstctlcite{IEEEexample:BSTcontrol}
    \title{iEmoTTS: Toward Robust Cross-Speaker Emotion Transfer and Control for Speech Synthesis based on Disentanglement between Prosody and Timbre}
    \author{Guangyan Zhang, Ying Qin, Wenjie Zhang, Jialun Wu, Mei Li, Yutao Gai, Feijun Jiang, Tan Lee\\
      \thanks{
      Guangyan Zhang and Tan Lee are with the DSP \& Speech Technology Laboratory, Department of Electronic Engineering, The Chinese University of Hong Kong, Hong Kong. Email: \{gyzhang,tanlee\}@link.cuhk.edu.hk \\
      Ying Qin is with Institute of Information Science, Beijing Jiaotong University, Beijing 100044, China. Email:yingqin@bjtu.edu.cn \\
      Wenjie Zhang, Jialun Wu, Mei Li, Yutao Gai and Feijun Jiang are with Intelligent Connectivity, Cloud \& Technology, Alibaba Group, Hangzhou 311121, China. Email: \{chenxi.zwj, Wjl205742, meizi.lm, yutao.gyt, feijun.jiangfj\}@alibaba-inc.com.
      }
      }

\markboth{Journal of \LaTeX\ Class Files,~Vol.~18, No.~9, September~2020}%
{How to Use the IEEEtran \LaTeX \ Templates}

\maketitle

\begin{abstract}
The capability of generating speech with a specific type of emotion is desired for many human-computer interaction applications. Cross-speaker emotion transfer is a common approach to generating emotional speech when speech data with emotion labels from target speakers is not available for model training. This paper presents a novel cross-speaker emotion transfer system named iEmoTTS. The system is composed of an emotion encoder, a prosody predictor, and a timbre encoder. The emotion encoder extracts the identity of emotion type and the respective emotion intensity from the mel-spectrogram of input speech. The emotion intensity is measured by the posterior probability that the input utterance carries that emotion. The prosody predictor is used to provide prosodic features for emotion transfer. The timbre encoder provides timbre-related information for the system. Unlike many other studies which focus on disentangling speaker and style factors of speech, the iEmoTTS is designed to achieve cross-speaker emotion transfer via disentanglement between prosody and timbre. Prosody is considered the primary carrier of emotion-related speech characteristics, and timbre accounts for the essential characteristics for speaker identification. Zero-shot emotion transfer, meaning that the speech of target speakers is not seen in model training, is also realized with iEmoTTS. Extensive experiments of subjective evaluation have been carried out. The results demonstrate the effectiveness of iEmoTTS compared with other recently proposed systems of cross-speaker emotion transfer. It is shown that iEmoTTS can produce speech with designated emotion types and controllable emotion intensity. With appropriate information bottleneck capacity, iEmoTTS is able to transfer emotional information to a new speaker effectively. Audio samples are publicly available\footnote{https://patrick-g-zhang.github.io/iemotts/}.
\end{abstract}


\IEEEpeerreviewmaketitle

\section{Introduction}

\IEEEPARstart{T}{he} basic function of a text-to-speech (TTS) system is to produce fluent speech from text input. Early approaches to TTS included waveform concatenation \cite{hunt1996unit,taylor2009text} and statistical parametric synthesis \cite{tokuda2000speech, zen2009statistical, qian2014training}. In recent years, TTS systems have been predominantly based on deep neural network models \cite{wang2017tacotron,shen2018natural,ren2020fastspeech, kim2021conditional}. State-of-the-art neural TTS systems are able to generate highly intelligible speech with human-level perceived naturalness. Nevertheless, the ability to realize expressiveness pertinent to natural speech communication is generally considered inadequate or absent. Expressiveness in natural speech is related to the speaker's affect status, namely emotion, mood and attitude, which depend on the communication scenario, and it is mainly manifested by prosody of speech \cite{zhang2021estimating,taylor2009text}. For applications like spoken dialogue systems, chat avatars, and video game dubbing, computer-generated speech is often desired to carry specific emotion types, e.g., anger, and fear, to create an engaging user experience. This research is about the realization of neural TTS systems for generating speech with different emotion types and controllable emotion intensity. 

Emotional speech generation with neural TTS models has been investigated extensively \cite{yamagishi2005acoustic,lee2017emotional,lorenzo2018investigating, um2020emotional}. A straightforward approach is to incorporate a categorical identity of emotion type as the auxiliary input to the TTS model to steer or control the characteristics of output speech. This approach requires training utterances with manually labelled emotion types. The preparation of such training data is known to be costly and time-consuming. To tackle the data sparsity problem, a common practice is fine-tuning a pre-trained model with a small number of labelled data \cite{tits2019exploring}. Semi-supervised learning was also attempted \cite{wu2019end}. In these approaches, manually labelled data for the target speakers is indispensable.

When reliably labelled data is absent, emotional speech can be generated via transferring emotion-related speech characteristics from source speakers to target speakers, defined as cross-speaker emotion transfer. In this process, we need utterances with emotion labels from the source speakers and a batch of utterances from the target speakers. To date, there has been little quantitative and detailed analysis of the emotional characteristics of the speech of the target speakers. Then the following question would be of our interest: \textit{does the speech of the target speakers carry any emotion?} 
In the present study, it is assumed that all emotion types are affectively valenced and cannot be affectively neutral \cite{ortony1990s}. One possible approach to generating emotional speech is to learn a `neutral to emotional' transformation from source speakers, which is applied to the neutral utterances of target speakers. There exist several problems with this approach. First, the neutral speech of target speakers may not be available, or the assumption of neutrality is not valid in the first place. 
Second, in the case that the neutral speech of source speakers is also unavailable, the `neutral to emotional' transformation cannot be learned,  making the cross-speaker emotion transfer not realizable. In this study, we do not need the assumption that the speech of target speakers must be neutral. Emotional characteristics in the speech of target speakers are to be learned from the manually annotated speech of source speakers in a semi-supervised manner.

The disentanglement of emotion (style) and speaker factors is an important issue in the realization of cross-speaker emotion transfer. Typically, the speaker factor is represented by an embedding, which can be derived from either a trainable look-up table \cite{gibiansky2017deep} or a pre-trained speaker verification system \cite{wan2018generalized,jia2018transfer}. The emotion embedding is extracted from reference speech using a reference encoder module \cite{skerry2018towards}. The learned emotion embedding may depend on other speech factors, e.g., speaker, text content, environment \cite{skerry2018towards,zhang2021estimating,zhang21u_interspeech,hsu2019disentangling}. Disentangling emotion from speaker identity would be a challenging problem when, for example, a type of emotion is carried by only one of the source speakers in the given dataset. This emotion type is essentially a speaker-specific one. As the speaker and emotion are highly entangled, separating them from each other is difficult. Instead, the presented study investigates disentanglement between timbre and prosody to address the problem of strong speaker-emotion entanglement. Emotion in speech is conveyed and mainly perceived via prosody \cite{zhang2021estimating, taylor2009text}. On the other hand, timbre is defined as the perceptual attribute that distinguishes two sounds with the same pitch, intensity, and duration\cite{american1973american}. Timbre in speech is closely related to the speaker's voice \cite{qian2020unsupervised}. Different speakers' voices can be distinguished based on timbre cues. According to the definition of timbre, there exists a complementary relationship between prosody and timbre; therefore, disentangling timbre information from prosody is expected to be more robust and generalized than disentanglement between speaker and emotion. We also note some recent works ~\cite{shechtman21_interspeech,pan2021cross,raitio20_interspeech} also leverage prosodic information to increase the controllability of style or emotion expression, while those models are not designed for the timbre-prosody disentanglement.

Emotion labels provided in speech datasets are often in the form of descriptive affect states, e.g., anger, happiness, and sadness. In reality, emotional expressions ingrained in human speech are nuanced with varying intensity. While obtaining categorical emotion labels is laborious and expensive, labelling the emotion intensity of speech utterances is even more challenging. There is no universal or standard model to describe emotion intensity in speech. Thus, poor agreement and consistency in intensity labels are expected. Recent studies attempted unsupervised labelling and control of emotion intensity in speech \cite{um2020emotional,zhu2019controlling, sorin2020principal, li2021controllable, wu2021cross,li2022cross}. In this paper, a new approach to probability-based emotion intensity learning and control is investigated to generate emotional speech with designated intensity levels, e.g., weak, moderate and high. We hypothesise that if one utterance with a specific emotion type can be distinguished easily from utterances carrying other emotion types by a speech emotion classification model, this utterance should have a higher perceived emotion intensity regarding this emotion and vice versa. The emotion intensity could be measured in terms of the posterior probability of that emotion type derived from this classification model. 

In zero-shot emotion cross-speaker transfer, emotion is transferred to the speech of target speakers who are unseen to the model in training. With zero-shot cross-speaker emotion transfer, we can generate the emotional speech for an unseen target speaker and do not need to train the system again. Zero-shot voice conversion was studied extensively, while little work has been done on cross-speaker emotion transfer. The information bottleneck mechanism is applied to improve the performance of zero-shot cross-speaker emotion transfer in this study.

This paper describes a novel system for cross-speaker emotion transfer. The system comprises three core modules: emotion encoder, prosody predictor, and timbre encoder. The emotion encoder is used to extract discrete emotion types and emotion intensity values from input utterances. Given the emotion type and intensity, the prosody predictor generates prosodic features in accordance with the input phoneme sequence and emotion type. The timbre encoder provides the timbre information of output speech. If speech data of target speakers is available, the timbre encoder is realized as a trainable layer for embedding lookup. In the case of zero-shot emotion transfer, the timbre encoder consists of a speaker encoder and a bottleneck layer. The ground-truth prosodic features are used for system training. The system can remove the prosodic information but retain timbre-related information for the timbre encoder. The main contributions of this study are summarized as follows:
 
 \begin{itemize}
  \item A novel design of the TTS system named iEmoTTS for cross-speaker emotion transfer is proposed. To our knowledge, previous methods of cross-speaker emotion transfer were commonly based on speaker-emotion disentanglement. In iEmoTTS, we develop and apply a timbre-prosody disentanglement based approach to cross-speaker emotion transfer. The proposed iEmoTTS demonstrates superior performance to other emotion transfer systems reported recently. In the ablation study, iEmoTTS is compared with its variants to demonstrate the effectiveness of timbre-prosody disentanglement;
  \item An emotion encoder is specifically designed for end-to-end semi-supervised training. The emotion encoder produces discrete emotion type IDs for speech data of target speakers in training and is trained with other components end-to-end; 
  \item A new method of probability-based emotion intensity control is developed and realized in the emotion encoder of iEmoTTS. With this method, synthesized speech with varying levels of emotion intensity can be generated by the iEmoTTS;
  \item  Zero-shot cross-speaker emotion transfer is achieved with iEmoTTS by a specifically designed timbre encoder with a bottleneck layer. Experimental results show that emotional information can be transferred to unseen target speakers while maintaining a certain degree of speaker similarity.
\end{itemize}

The remaining part of this paper proceeds as follows. \Cref{related_work} introduces the related work. \Cref{analysis_emotion} analyzes the emotional characteristics in the speech. The overview and each component of the proposed iEmoTTS are described in \Cref{iemotts_overview}. \Cref{experiments} introduces the datasets, configurations and evaluation metrics for the experiments. \Cref{experiments_results} presents the experiment results of the research.

\section{Related works}
\label{related_work}

\subsection{Cross-speaker emotion transfer}
\label{cross_speaker_emotion_transfer}
Cross-speaker emotion or style transfer has been studied in the era of statistic parametric speech synthesis \cite{lorenzo2013towards, ohtani2015emotional, inoue2017investigation, chen2014speaker, hodari2018learning} and lately with neural TTS models \cite{whitehill20_interspeech, gao20e_interspeech, cai2021emotion, wu2021cross, bian2019multi, shechtman21_interspeech, lu2021multi, pan2021cross, shin22b_interspeech}. Existing approaches can be divided into two main categories. Methods of the first category, e.g., \cite{lorenzo2013towards, ohtani2015emotional, inoue2017investigation, chen2014speaker, whitehill20_interspeech, bian2019multi, lu2021multi, pan2021cross, shechtman21_interspeech, xue2021cycle}, require or assume that the given speech of target speakers are neutral. Emotion transfer can be carried out using a `neutral to emotional' transformation learned from the speech of source speakers. Lorenzo-Trueba et al. \cite{lorenzo2013towards} proposed an emotion transplantation method in the HMM-based speech synthesis framework. Two linear transformations were applied to transfer the acoustic features of neutral speech to the feature spaces of the target speaker and style, respectively. Whitehill et al. \cite{whitehill20_interspeech} applied the principle of adversarial cycle consistency \cite{zhu2017unpaired} to encourage the synthesized speech to preserve the appropriate styles. Recently, Pan et al. \cite{pan2021cross} used a prosody bottleneck layer to predict prosodic features for cross-speaker emotion transfer. The prosodic features are predicted at the inference stage given source speakers' speaker and emotion type IDs. The predicted prosodic features are combined with speaker identity information to generate the speech. This approach can be seen as copying the predicted prosody of source speakers and applying it to the target speakers. Lu et al. \cite{lu2021multi} attempted to address the problem of cross-speaker emotion transfer by introducing phoneme-level latent features. These fine-grained features disentangled from the speaker, tone, and global emotion information are used to condition the backbone TTS system. Shechtman et al.\cite{shechtman21_interspeech} proposes a controllable style transfer with prosodic descriptors as input, which allows the user to control the expression strength for each expressive style.

In the second category of approaches, the speech utterances of target speakers are not presumed to be neutral. Instead, they are labelled using a pre-trained speech emotion classifier. The classifier can be trained independently using manually labelled speech data \cite{gao20e_interspeech, cai2021emotion, hodari2018learning}. It can also be included as part of the emotion transfer TTS system and jointly trained with the speech of target speakers in a semi-supervised manner \cite{wu2021cross}. Gao et al. \cite{gao20e_interspeech} used a pre-trained emotion classifier to label the speech of target speakers and trained the emotional TTS with these labelled data. A joint-training approach was proposed in \cite{wu2021cross}. A set of emotion tokens were defined, and each was mapped to an emotion type by a classifier. The emotion embeddings for both source and target speakers were derived via attention between the prosody embeddings\cite{skerry2018towards} and emotion tokens. At the emotion transfer stage, the emotion embedding was determined from emotion tokens given emotion type ID. The potential problem is that emotion embeddings generated in the training and transfer stage might be inconsistent.

\subsection{Speech factors disentanglement}
Speech factor disentanglement has been studied in various tasks of speech technology, namely automatic speech recognition, TTS \cite{skerry2018towards,wang2018style, an2022disentangling } and voice conversion \cite{qian2019autovc,qian2020unsupervised}. The relevant methods were also investigated in emotion, style and prosody transfer. Skerry-Ryan et al. proposed to disentangle a prosody representation from speech content and speaker characteristics and hence achieve prosody transfer in the Tacotron-based TTS model \cite{skerry2018towards}. The prosody representation is extracted from mel-spectrograms of speech. It is assumed to retain the pertinent prosodic information that can be used in conjunction with the text transcript and speaker information to reconstruct input speech. 
For cross-speaker emotion (style) transfer, a multi-reference TTS stylization system based on GST-Tactotron was presented in \cite{bian2019multi}. An intercross training scheme was also proposed, in which the different reference encoder is used to decompose and control a specific style class. Speaker and emotion are considered two different style classes. Whitehill et al. \cite{whitehill20_interspeech} proposed a method based on adversarial cycle consistency to ensure all possible style combinations were used. Being designed for speaker and style level disentanglement, these methods might meet the following challenges: 1) certain types of emotion may be tightly entangled with specific source speakers; 2) the acoustic conditions from which speech of source speakers and target speakers are acquired could be very different in practice. Qian et al.\cite{qian2019autovc} developed a voice conversion system named AutoVC that disentangles speech content from timbre information. Furthermore, a voice conversion model \cite{qian2020unsupervised} was introduced to decompose speech into four components: pitch, rhythm, speech content, and timbre. 


\subsection{Emotion Intensity Control}
Variation of emotion intensity (strength) in synthesized speech can be achieved via a learned emotion representation (embedding). Straightforwardly, this can be done by applying a scalar weight to the emotion embedding \cite{li2021controllable, wu2021cross} at the inference stage. This approach was shown effective in image style transfer\cite{Johnson2016Perceptual}. Zhu et al. \cite{zhu2019controlling,leiyi2021finegrain} proposed to quantize emotion intensity by learning a rank function using the concept of relative attribute \cite{parikh2011relative}. In order to learn the ranking function, prosodic features were extracted from  $<$neutral, emotional$>$ utterance pairs. The emotion intensity values were measured by weighting the prosodic features with learnable ranking weights and subsequently applied as an auxiliary input to train the TTS model. In \cite{um2020emotional}, an emotion intensity control method was proposed to control the distinct characteristic of a target emotion category. An interpolation technique was introduced to control emotion intensity by gradually changing the emotional to neutral speech. Two challenges need to be overcome in these approaches: 1) the emotion intensity has to be derived using a separate ranking function \cite{zhu2019controlling, leiyi2021finegrain} or by prosody analysis \cite{um2020emotional,sorin2020principal}; 2) $<$neutral, emotional$>$ utterance pairs are required. 

\section{ Analysis of emotion-related characteristics in speech}
\label{analysis_emotion}
\subsection{Emotion Speech Corpora for Analysis}
Two corpora of Chinese speech \footnote{All corpora used in this work are from https://www.data-baker.com/.} are used for analysing emotion-related speech characteristics. One of the corpora is with human-annotated emotion labels (\textbf{Multi-S60-E3}), and the other is unlabeled (\textbf{VA-S2}).

The \textbf{Multi-S60-E3} corpus contains 30,000 speech utterances from 30 male and 30 female speakers. The utterances were elicited to carry three different emotion types: anger, happiness, sadness, and neutrality. A total of 125 sentences are combined with different speakers and emotions, so speech content is independent of speaker and emotion type. The \textbf{VA-S2} corpus was created for building TTS systems on voice assistant devices. It contains one female (\textbf{F1}) and one male (\textbf{M1}) speaker, each having about 2,000 utterances.

\subsection{Analysis of Characteristics of Emotional Speech}
\label{emotion_classifier}
A speech emotion recognition (SER) model has been trained on the \textbf{Multi-S60-E3} dataset. The model consists of a feature extractor that follows the CNN+RNN structure as in \cite{satt17_interspeech,skerry2018towards} and a softmax layer. The feature extractor takes mel-spectrogram of an utterance as input and processes it in five Conv-Norm layers\cite{skerry2018towards}. The output is then passed to a bidirectional GRU layer. Only the hidden state of the last time step is selected and projected to the hidden feature and then four-dimensional logits, also known as unnormalized log probabilities. The four-class posterior probability distribution is obtained by passing the logits to the softmax layer. The accuracy of SER on the test set is 97.11\%.

\begin{figure}[!t]
  \centering
  \includegraphics[width=\linewidth]{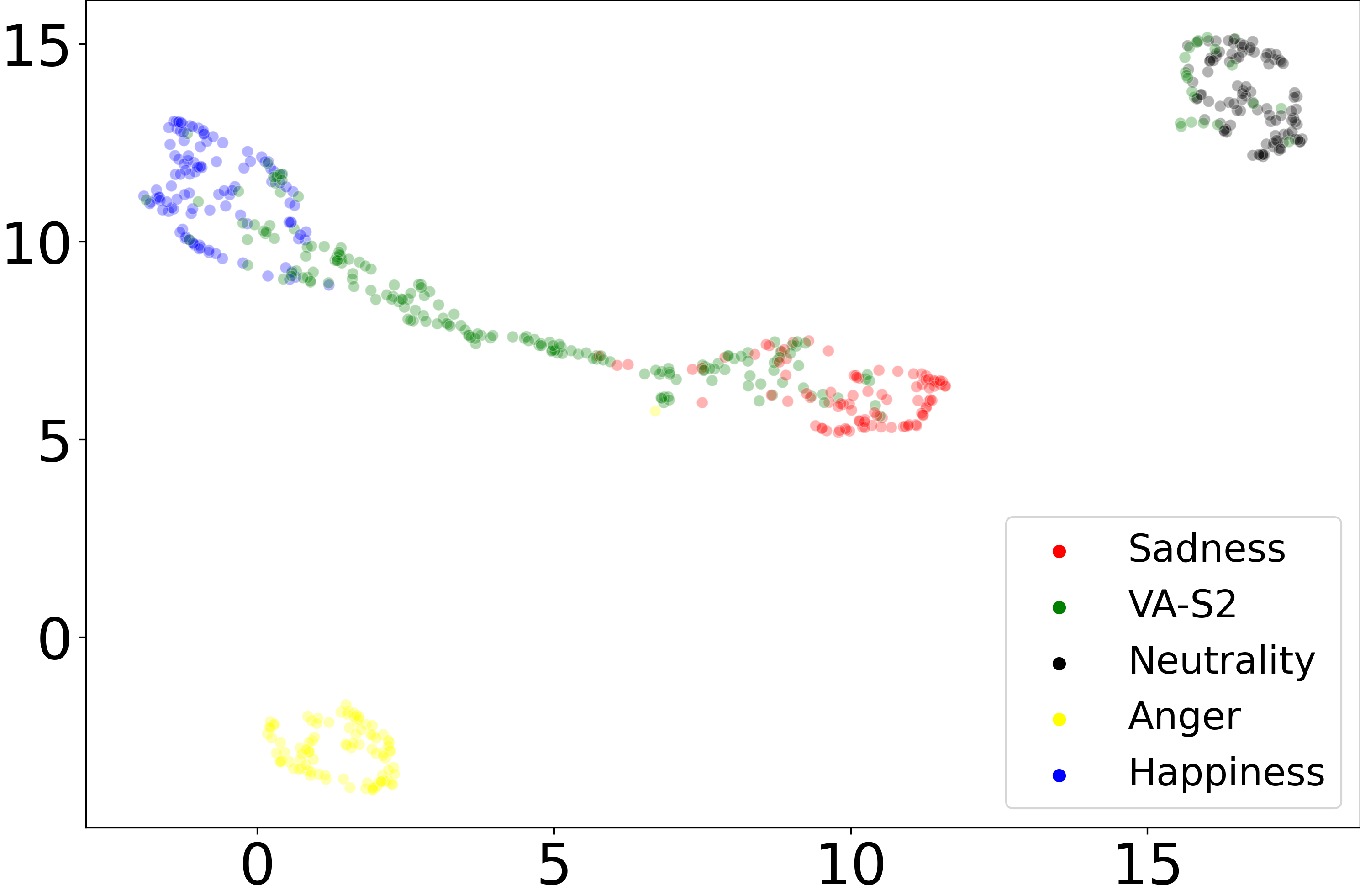}
  \caption{The visualization of the hidden features for utterances randomly selected from the \textbf{VA-S2} and \textbf{Multi-S60-E3}. The corresponding utterances for the green points are from \textbf{VA-S2}. The points in red, black, yellow, and blue colours are from \textbf{Multi-S60-E3} and represent sadness, neutrality, anger and happiness, respectively.}
  \label{Figure:umap_ill}
\end{figure}

The SER model predicts the emotion types of all utterances in \textbf{VA-S2}. The prediction results are shown in \Cref{pred_table}. It is noted that most utterances in \textbf{VA-S2} are classified as being non-neutral. If speakers in \textbf{VA-S2} are used as target speakers, it would not be appropriate to assume them to be neutral like some cross-speaker emotion transfer methods. 400 utterances are randomly selected from \textbf{Multi-S60-E3} and 200 utterances from \textbf{VA-S2} corpus. The hidden features of these utterances are visualized using the UMAP\cite{mcinnes2018umap} as shown in \autoref{Figure:umap_ill}. The utterances from \textbf{Multi-S60-E3} are clustered into four clusters, each corresponding to one of the three emotion types or neutrality. The utterances from \textbf{VA-S2} exhibit greater distances from the centres of the four clusters of \textbf{Multi-S60-E3}. From our informal listening test, the utterances of \textbf{VA-S2} are perceptually consistent with the predicted emotion types but seemingly with lower intensity of emotion expression. In short, it would be more appropriate to represent each utterance in \textbf{VA-S2} with an emotion type and an intensity value.

\begin{table}[htbp]
        \caption{The Prediction results of passing the speech utterances in \textbf{VA-S2} corpus to a trained SER model.  }
    \centering
    \begin{tabular}{m{1.2cm}m{1.2cm}m{1.2cm}m{1.2cm}m{1.2cm}}
         \toprule 
           &  Sadness &  Neutrality & Happiness &  Anger   \\
         \midrule
         \textbf{M1} & 4.15\%  &  19.83\% & \textbf{73.07\%} & 2.95\% \\
         \textbf{F1} & 23.59\% &  16.38\% & \textbf{56.86\%} & 3.17\% \\
         \bottomrule
    \end{tabular}
    \label{pred_table}
\end{table}

\begin{figure}[!t]
  \centering
  \includegraphics[width=\linewidth]{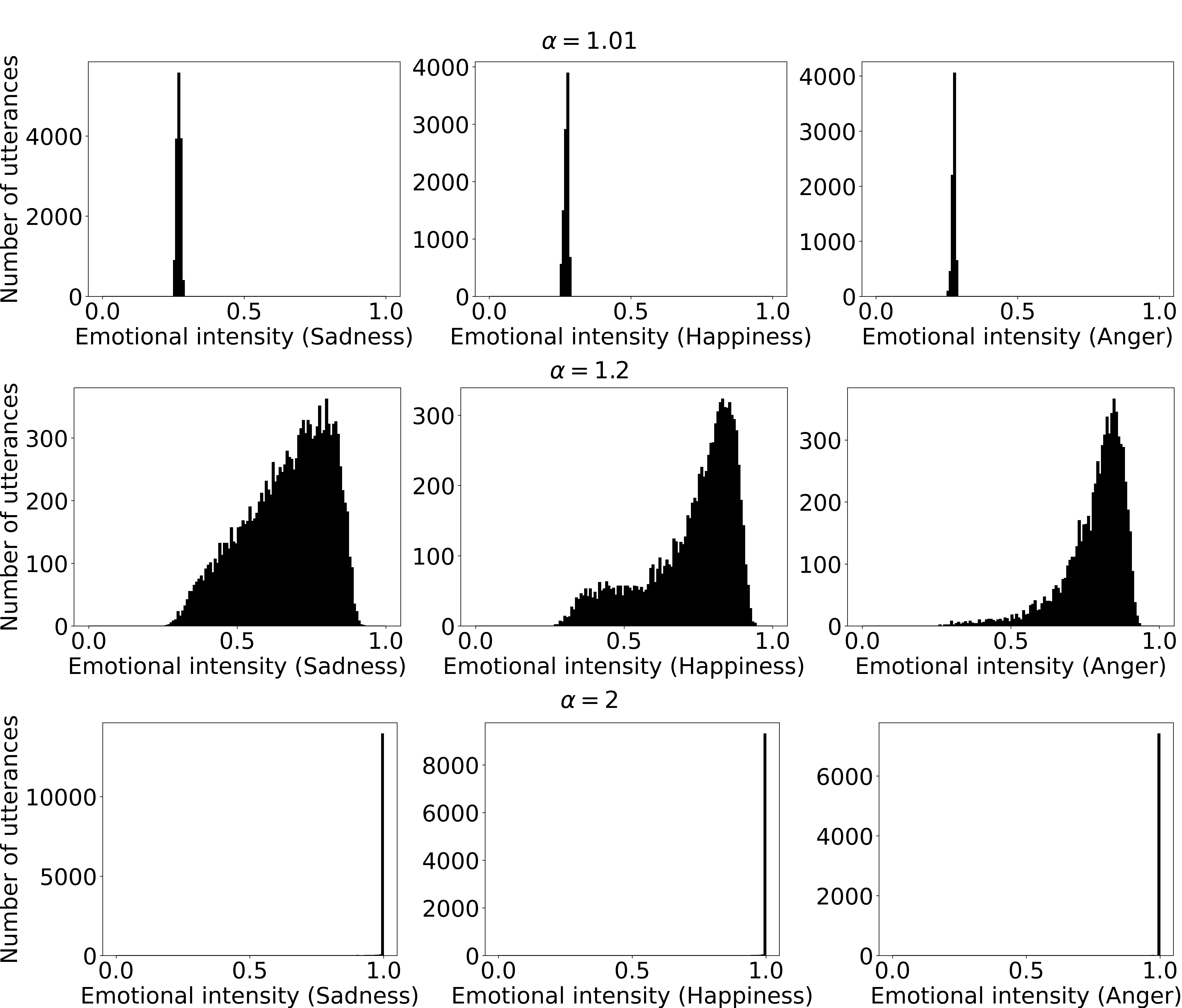}
  \caption{The histogram of $int_i$ for the emotion types, happiness, sadness and anger when $\alpha$ is 1.01,1.2 and 2.}
  \label{Figure:hisgram_sum}
\end{figure}

We propose to measure the emotion intensity of a speech utterance in terms of the posterior probability. The posterior distribution $\boldsymbol{\pi}$, which is given as the output of the SER model, is obtained by normalizing the logits $\mathbf{z}$ with a softmax function. Specifically, for $N$ distinct emotion types, the probability for emotion type $i$ given logits $\mathbf{z}$ is computed as,
\begin{align}
  \boldsymbol{\pi}_i = \text{softmax}(\mathbf{z})_i = \frac{\exp(\mathbf{z}_i)}{\sum_{j=1}^{N} \exp(\mathbf{z}_j)}
\end{align}
Human-annotated emotion labels are used for the SER model training with the cross-entropy loss function. The optimizer maximizes log-likelihood $\log \boldsymbol{\pi}_i$, which pushes up $\mathbf{z}_i$, and thus reduces the value of $\mathbf{z}_{j  \neq i}$. For inference, the predicted emotion type of input utterance is used. The degree of value $\mathbf{z}_i$ over other logits values $\mathbf{z}_{j  \neq i}$ can be regarded as the classifier's confidence in the prediction. The confidence level might be related to the emotion intensity, where a higher confidence level indicates stronger emotion intensity and vice versa. The logits $\mathbf{z}$ need to be normalized across different utterances. The softmax function is a usual choice to normalize $\mathbf{z}$ as a categorical distribution. Noting that and $\text{softmax} ((\mathbf{z})_i)$ may be saturated to 1, the softmax function is modified such that the base $\mathbf{e}$ is changed to a hyper-parameter $\mathbf{\alpha}$. The posterior probability-based intensity $int_i$ for the emotion type $i$ is defined as,
\begin{equation}
    int_i = \frac{\alpha^{\mathbf{z}_i}}{\sum_{j=1}^{M} \alpha^{\mathbf{z}_j}} \\
\end{equation}
Different values of $\mathbf{\alpha}$, e.g., $1.01$, $1.2$, $2$ were experimented with the \textbf{VA-S2} corpus. The histograms of $int_i$ obtained on different emotion types are shown in \autoref{Figure:hisgram_sum}. When $\mathbf{\alpha}$ is close to 1, $int_i$ approaches $1/N$, where $N=4$ in our experiment. If $\mathbf{\alpha}$ is large, e.g., 2, $int_i$ would be saturated to 1. By choosing the appropriate value of $\mathbf{\alpha}$, e.g.,$1.2$, $int_i$ can have values in the entire interval from 0 to 1

  \begin{figure*}[t]
  \centering
  \includegraphics[width=0.9\textwidth]{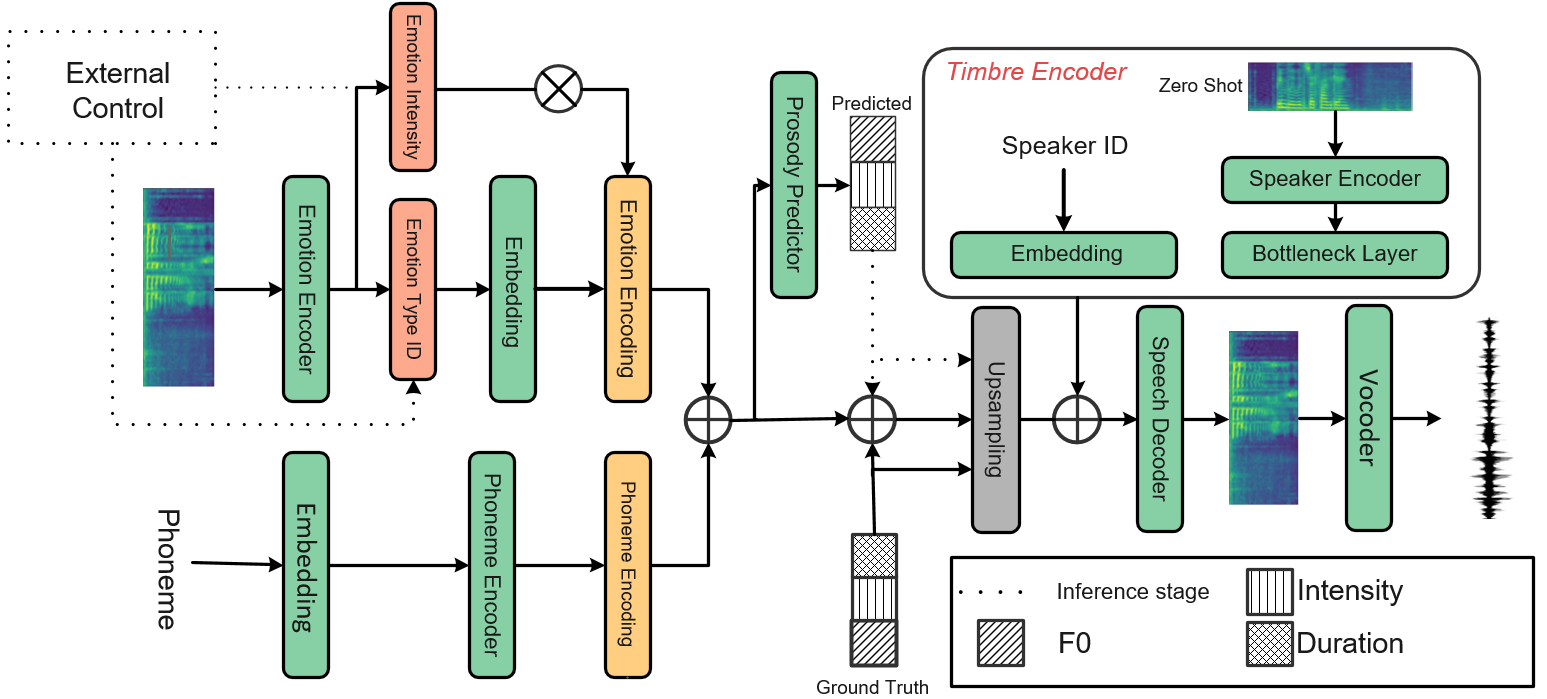}
  \caption{The architecture of proposed iEmoTTS model.}
  \label{Figure:model_arch}
\end{figure*}

\section{The Proposed iEmoTTS System}
\label{iemotts_overview}
\autoref{Figure:model_arch} illustrates the proposed iEmoTTS system. It consists of six major components: phoneme encoder, emotion encoder, prosody predictor, timbre encoder, speech decoder and vocoder. In training, iEmoTTS operates as a conditional auto-encoder, which aims to reconstruct the mel-spectrogram of an input speech utterance, given the phoneme sequence and speaker identity. This section first gives an overview of the system. Details about the component modules are presented separately.

\subsection{Model Overview}
\label{iemotts_model_overview}
A trainable lookup embedding layer is used to encode the phoneme sequence into a sequence of embedding, which is then encoded into a phoneme encoding sequence by the phoneme encoder. The emotion type ID is passed through another lookup embedding layer to produce an embedding, multiplied by the emotion intensity value to produce the emotion encoding. The emotion encoding is broadcasted and then added to the phoneme encoding sequence, producing a hidden sequence with dimension size 256 as the input of the prosody predictor. The 3-dimensional phoneme-level prosodic features are transformed into a sequence of 256-dimensional vectors by a 1-D convolutional layer, which is then added to the hidden sequence and up-sampled to frame-level representations. The timbre encoding generated by the timbre encoder is broadcast-added with the above frame-level representations and presented to the speech decoder to generate mel-spectrograms. A pre-trained vocoder converts the predicted mel-spectrograms to the speech waveform.

The phoneme encoder and speech decoder remain the same as the standard FastSpeech, using the feed-forward Transformer (FFT) \cite{ren2019fastspeech} block as the basic unit. During modelling training, the emotion type ID and intensity are derived from the emotion encoder given input mel-spectrograms. For emotion transfer (inference), the emotion type ID and intensity are provided to the model by external control. 

The iEmoTTS is designed to perform disentanglement between timbre and prosody for the cross-speaker emotion transfer task. In this design, prosody is acoustically represented by pitch, intensity and duration~\cite{karlapati20_interspeech}. Disentanglement of prosody and timbre is achieved by (1) incorporating ground-truth prosodic features in model training; and (2) presenting the timbre encoding to the model after the prosodic features are incorporated. The timbre encoding is derived from the timbre encoder with speaker ID or mel-spectrogram as input which includes both timbre and prosodic information of speakers. However, the prosodic information is provided first to the model, and it is assumed there exists a complementary relationship between timbre and prosody. From an information-theoretic perspective, the iEmoTTS should be able to get rid of prosodic information from the timbre encoding but retain the timbre information. 
 \begin{figure}[!t]
  \centering
  \includegraphics[width=\linewidth]{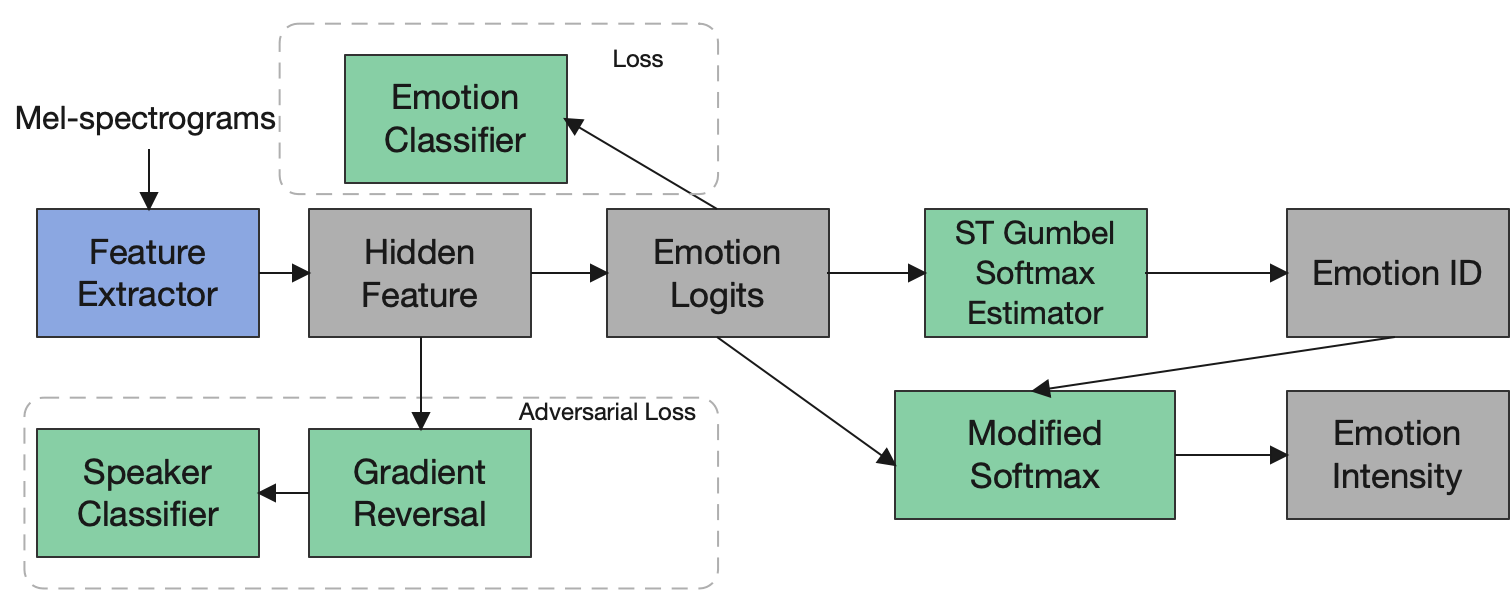}
  \caption{The overview of the emotion encoder.}
  \label{Figure:emotion_enc}
\end{figure}

\subsection{Emotion Encoder}
\label{emotion_encoder_desc}
The emotion encoder determines the emotion type ID and the respective posterior probability-based emotion intensity values from the mel-spectrogram. Specifically, the emotion encoder can produce discrete emotion type ID consistent with the external control input in the inference stage while enjoying an end-to-end training pipeline. As illustrated in \autoref{Figure:emotion_enc}, the emotion encoder is composed of a feature extractor, an adversarial speaker classifier, an emotion classifier, a modified softmax layer and a Straight-Through (ST) Gumbel-Softmax estimator.

\subsubsection{Semi-supervised Training}
Note that only utterances from source speakers have emotion labels. Suppose a total $N$ emotion types, including neutrality, are represented in the speech data of source speakers.
The feature extractor converts the input mel-spectrograms from both source and target speakers into hidden features, as described in \Cref{emotion_classifier}. Speaker-related information in the hidden features is partly suppressed by passing the hidden features to a gradient reversal layer \cite{ganin2016domain} and a speaker classifier which consists of a dense and softmax layer. Each hidden feature is then projected onto $M$-dimensional logits $\mathbf{z}$ by one fully connected layer. Only the logits belonging to source speakers (with emotion labels) will be passed to the emotion classifier. The emotion classifier comprises a softmax layer, which forces the first $N$ dimensions of  $\mathbf{z}$ to correspond to the $N$ emotion types. Note that the $M \ge N$ in case some utterances from target speakers may not carry any known emotion types from source speakers. The ST Gumbel-Softmax estimator sampled the emotion type ID $i$ from the logits $\mathbf{z}$. The emotion intensity scalar is obtained as the $i_{th}$ value from the result of applying the modified softmax function to $\mathbf{z}$.

\subsubsection{Straight-Through Gumbel-Softmax Estimator}
The emotion type $i$ is a discrete quantity obtained by sampling from a categorical distribution. This sampling process is not differentiable; thus, gradient back-propagation is not applicable. The Gumbel-Softmax distribution \cite{jang2017categorical, maddison2017concrete} can be used to approximate sampling from a categorical distribution $\boldsymbol{\pi}$:
\begin{align}
  \mathbf{y}_i = \frac{\exp((\log(\boldsymbol{\pi}_i) + \mathbf{g}_i)/\tau)}{\sum_{j=1}^M \exp((\log(\boldsymbol{\pi}_j) + \mathbf{g}_j)/\tau)}, i = 1,2,...,M
\end{align}
where $\mathbf{g}_i$ are i.i.d samples drawn from Gumbel(0,1) and $\mathbf{y}$ denotes a sample from the Gumbel-Softmax distribution. $\tau$ is the softmax temperature. $\mathbf{y}$ becomes one-hot if $\tau$ approaches 0. In practice, we start at a high temperature and anneal to a small but non-zero temperature. Also, $\log \boldsymbol{\pi}_i$ is approximated by $\mathbf{z}_i$. $\mathbf{y}$ is converted into one-hot discrete data $\mathbf{y'}$ using $\arg \max$ operation to represent the emotion type. To ensure that the back-propagation can proceed, the gradient of $\mathbf{y'}$ in the backward pass is used to approximate the gradient of $\mathbf{y}$, $\Delta_\theta \mathbf{y} \approx \Delta_\theta \mathbf{y'}$.

\subsection{Prosody Predictor}
The prosody predictor aims to predict F0, intensity and duration features for cross-speaker emotion transfer. The prosody predictor has a similar structure to the variance adaptor in \cite{ren2020fastspeech} but with a larger model size.
It consists of a six-layer 1D-convolutional network with ReLU activation, followed by layer normalization, a dropout layer and an extra linear layer.
In addition to model size, the proposed prosody predictor differs from the variance adaptor in several aspects. First, the variance adaptor consists of a sequence of independent predictors, each for a specific prosodic feature. The proposed predictor outputs a multiple-dimension vector in which F0, duration and intensity are represented by different dimensions. This design may help the model to capture the potential dependence among different prosodic features. Second, the proposed predictor generates prosodic features at the phoneme level instead of the frame level. Phoneme-level F0 and intensity values are obtained by averaging the frame-level values over the time intervals of the respective phonemes.
On the contrary, the variance predictor outputs frame-level pitch and energy. Frame-level features are challenging to predict and would affect the model generalization capacity\cite{zhangstudy222}. Lastly, prosodic features in iEmoTTS are mean-variance-normalized on individual speakers.

\subsection{Timbre Encoder}
The role of the timbre encoder is to provide timbre-related information to the TTS model. For the cross-speaker emotion transfer task where the target speakers are seen in model training, a trainable lookup embedding layer with speaker ID as input is used to produce the timbre encoding. 

For zero-shot cross-speaker emotion transfer, a pre-trained speech encoder is employed to extract a speaker embedding \cite{jia2018transfer} from the mel-spectrograms. The speech embedding is then passed through a bottleneck layer to generate timbre encoding. The pre-trained speaker encoder follows the design in \cite{wan2018generalized}, which consists of a stack of two LSTM layers. The last time output of the LSTM layers is selected and projected down to a speaker embedding with a fully connected layer. The speaker embedding might still contain redundant information regarding prosody, content or channel noise. A bottleneck layer is leveraged to remove the redundant information and keep most of the timbre information. The bottleneck layer takes the speaker embedding as the input and outputs timbre encoding. This paper implements the bottleneck layer by a modified Vector Quantized VAE (VQ-VAE) quantized layer \cite{van2017neural,zhang21u_interspeech}. It learns a dictionary (codebook) $\mathbf{E}$ with dictionary size as $K$ and group number $G$. The speaker embedding is first divided equally into $G$ groups and arranged as a matrix, where each column can be encoded by an integer index independently. Each index will then query a corresponding embedding from the dictionary $\mathbf{E}$, and all corresponding embeddings will concatenate together as the timbre encoding. With a fixed dictionary size, the information bottleneck (IB) capacity of the bottleneck layer (i.e., $G \log K$) can be controlled by changing the group number $G$. The larger the number $G$, the higher the IB capacity, and the more information can pass through the bottleneck layer. This study sets the dictionary size $K$ and group number $G$ as 32 and 4, respectively. In training, timbre encoding is different for each sentence. At the inference stage, the timbre encoding for each speaker is generated by feeding one or several utterance(s) of the same speaker to the timbre encoder and averaging the resulting sentence-level encoding.

\subsection{Objective Function}
The loss function of iEmoTTS consists of four components,
\begin{align}
  L = L_{mel} + \lambda_1 L_{pros} + \lambda_2 L_{adv\_spk} + \lambda_3 L_{emo\_source} 
\end{align}
where $L_{mel}$ is the Mean Absolute Error (MAE) between the synthesized mel-spectrograms and the ground truth, $L_{pros}$ is the L2 loss between the predicted and ground-truth prosodic features, and $L_{adv\_spk}$ and  $L_{emo\_source}$ denote the cross-entropy losses for the adversarial speaker classifier and the emotion classifier, respectively. $\lambda_1$, $\lambda_2$ and $\lambda_3$ are the hyper-parameters to balance the contributions of different losses. Since emotion and speaker are highly entangled in training data, a very small weight $\lambda_2$ is assigned to the loss function of the adversarial classifier $L_{adv\_spk}$ to prevent emotion information from being reduced. The $\lambda_1$, $\lambda_2$ and $\lambda_3$ are set as 0.8, 0.01 and 0.5, respectively.

\section{Experimental Protocol}
\label{experiments}
\subsection{Data Preparation}

Four speech corpora of Mandarin speech are used in the experiments on emotion transfer. They are \textbf{Multi-S60-E3}, \textbf{Child-S1-E6}, \textbf{VA-S2} and \textbf{Read-S40}. The emotion labels of \textbf{Multi-S60-E3} and \textbf{Child-S1-E6} are available. \textbf{Multi-S60-E3} is a multi-speaker emotional corpus as described in \autoref{analysis_emotion}. The \textbf{Child-S1-E6} corpus contains 12,000 speech utterances by a female voice actor imitating a child-like voice. There are six emotion types: happiness, amazement, anger, disgust, poorness, and fear. Each type has 2000 spoken utterances. The corpora \textbf{VA-S2} and \textbf{Read-S40} are without emotion labels. \textbf{VA-S2} is described in \autoref{analysis_emotion}. The other one, named \textbf{Read-S40}, contains 20 male and 20 female speakers, each having about 500 utterances in reading style.


\subsubsection{Speech data of source speakers}
The dataset \textbf{Child-S1-E6} covers six different emotion types. This corpus would be included as part of the training data as one source speaker in all experiments below. For cross-speaker emotion transfer where target speakers are seen in training, we involve additionally three female and three male speakers from  \textbf{Multi-S60-E3} as source speakers, and hence there are seven source speakers and seven different emotion types covered. While for zero-shot emotion transfer, in addition to \textbf{Child-S1-E6}, 15 female and 15 male speakers from \textbf{Multi-S60-E3} are used as source speakers to improve model generalizability \cite{qian2019autovc}. As a result, there are 31 source speakers with seven emotion types in the experiment on zero-shot cross-speaker emotion transfer. It should be noted that some of the emotion types, namely, amazement, disgust, fear, and poorness, are associated with source speakers only from \textbf{Child-S1-E6}. 

\subsubsection{Speech data of target speakers}
For the cross-speaker emotion transfer task, the two speakers from corpus \textbf{VA-S2} are used as target speakers. The target speakers are not seen in training in zero-shot emotion transfer. In this case, one female and one male speaker are randomly selected as target speakers from \textbf{Read-S40}. Only five utterances (around 20s) are used for each speaker.

\subsection{Implementation Details}
The raw Chinese text is converted into phoneme sequence by an open-sourced grapheme-to-phoneme tool\footnote{https://github.com/mozillazg/python-pinyin}. The Montreal forced alignment (MFA) tool \footnote{https://github.com/MontrealCorpusTools/Montreal-Forced-Aligner} is employed to obtain the phoneme boundaries and duration of the speech. 80-band mel-spectrograms are computed from raw speech waveforms with a frame length of 50ms and frameshift of 12.5ms. A pre-trained vocoder based on full-band HiFi-Gan \cite{kong2020hifi} is adopted to transform the predicted mel-spectrograms to a speech waveform.

The $N$ and $M$ in \Cref{emotion_encoder_desc} are set as 8 and 10. The iEmoTTS is trained with a batch size of 32 sentences on 4 NVIDIA V100 GPUs, using the Adam optimizer \cite{kingma2014adam} and the learning rate schedule in \cite{vaswani2017attention}.  It takes 200k steps for training until convergence.

\subsection{Subjective Evaluation Methods}
Crowdsourced Mean Opinion Score (MOS) is used for subjective evaluation of the aspects of emotion similarity, speaker similarity and voice quality of synthesized speech. The scores range from 1 to 5 in 0.5 point increments \cite{loizou2011speech}. A test of emotion intensity ranking is also performed. 

\textbf{Emotion Similarity}: Emotion similarity is most critical in the cross-speaker emotion transfer task. It measures how well the synthesized speech resembles designated emotion types. Subjective evaluation of emotion similarity starts by arranging for each listener to listen to 5-10 sample utterances of each emotion type from source speakers to let the listener understand and be familiar with the type. Subsequently, the listeners are asked to evaluate each audio sample based on how much it sounds like a given emotion type. The listeners are advised to ignore the contents of the utterance when scoring the emotion similarity. Listeners can move back to listen to the audio samples during the evaluation.

\textbf{Speaker Similarity}: The speaker similarity test evaluates whether the synthesized utterances carry the speaker characteristics of target speakers. Before evaluating the test samples of each target speaker, listeners have to listen to 10 utterances to create an overall impression of this speaker. Then, listeners are asked to evaluate each audio sample based on how much it resembles this target speaker.

\textbf{Voice Quality}: For voice quality assessment, listeners are presented with one utterance each time and asked to give a score regarding speech naturalness and pronunciation correctness. 

The evaluation of emotion intensity control is carried out via a ranking test \cite{li2022cross}. In each test trial, the listener is presented with three utterances synthesized with the same text content and high, moderate and low emotion intensity, respectively. The three utterances are presented in random order, and the listener is asked to classify them as high, moderate and low emotion intensity. A good classification accuracy implies that control of emotion intensity is effective.

In the following sections, the results of each evaluation metric shown in each table are from an independent subjective test. In a subjective test, 30 native Chinese speakers are presented with synthesized audio samples from different systems but with the same text content each time. We choose 20 test utterances per emotion category for a total of 140 for each system. Paired t-test is applied to verify that performance differences among the systems are significant in all MOS experiments. All reported significant differences are for $p \leq 0.05$.

\section{Experimental Results}
\label{experiments_results}

\subsection{Performance Evaluation}
Two recent cross-speaker emotion transfer models, Trans-CLN\cite{wu2021cross} and Trans-Pros\cite{pan2021cross} are used for performance comparison. Trans-CLN uses a set of emotion tokens to represent different types of emotions. It adopts a semi-supervised training strategy to map the utterances of target speakers to the emotion tokens. Trans-Pros is built upon the Transformer TTS \cite{li2019neural}. It predicts the bottle-necked prosodic features given the information of source speakers. The bottle-necked prosodic features are combined with the target speaker ID and text to generate the mel-spectrograms. Two reference systems and iEmoTTS share similar model parameters.

The compared results are shown in \autoref{cmp_other_methods}. In most cases, iEmoTTS performs better than the two reference systems. On emotion similarity, iEmoTTS and Trans-Pros show comparable performance, and both are significantly better than Trans-CLN. Trans-CLN does not achieve good emotion similarity, especially on the emotion types with low arousal levels, e.g., sadness and poorness. iEmoTTS and Trans-CLN show similar performance in speaker similarity, and Trans-Pros performs the worst. In voice quality, iEmoTTS performs significantly better than Trans-Pros and Trans-CLN. Two reasons might cause the unsatisfying performance of Trans-Pro: (1) The prosodic features are predicted from one source speaker and, therefore, might retain speaker-specific information from that speaker. (2) The speaker embedding of the target speaker might also retain the prosodic information of the target speaker.

\begin{table*}[htbp]
        \caption{Subjective evaluation results between proposed iEmoTTS and two recent works on cross-speaker emotion transfer. The Emotion similarity, speaker similarity and voice quality MOS with a 95\% confidence interval are compared. A higher MOS value indicates better performance. The MOS values significantly higher than other models are in bold.} 
    \centering
    \scalebox{0.92}{
    \begin{tabular}{lccc|ccc|ccc}
             \toprule
            \multirow{2}*{\textbf{Emotion}} & \multicolumn{3}{c}{Emotion Similarity MOS} & \multicolumn{3}{c}{Speaker Similarity MOS} & \multicolumn{3}{c}{Voice Quality MOS} \\
            \cmidrule(lr){2-4}\cmidrule(lr){5-7}\cmidrule(lr){8-10}
                       & Trans-CLN        & Trans-Pros                & iEmoTTS    & Trans-CLN        & Trans-Pros                & iEmoTTS & Trans-CLN        & Trans-Pros                & iEmoTTS \\
            \midrule

         Happiness     & $4.32 \pm 0.07$          & $\mathbf{4.53 \pm 0.08}$ & $4.38 \pm 0.06$  
                       & $\mathbf{4.00 \pm 0.10}$ & $3.50 \pm 0.14$          & $\mathbf{4.03 \pm 0.10}$ 
                       & $ 4.23 \pm 0.06$         & $ 3.95 \pm 0.08  $       & $\mathbf{4.36 \pm 0.08} $ \\
         Sadness       & $2.59 \pm 0.08$          & $\mathbf{4.32 \pm 0.07}$ & $3.98 \pm 0.08$ 
                       & $3.60 \pm 0.14$          & $3.12 \pm 0.12$          & $\mathbf{3.79 \pm 0.14}$ 
                       & $4.16 \pm 0.06$          & $3.70 \pm 0.09$          & $\mathbf{4.22 \pm 0.08}$  \\
         Poorness      & $3.05 \pm 0.10$          & $4.12 \pm 0.09$          & $\mathbf{4.30 \pm 0.07}$
                       & $3.74 \pm 0.09$          & $2.00 \pm 0.16$          & $\mathbf{3.87 \pm 0.17}$ 
                       & $\mathbf{4.15 \pm 0.07}$ & $3.33 \pm 0.10$          & $3.70 \pm 0.09$              \\
         Fear          & $3.22 \pm 0.09$          & $4.39 \pm 0.08$          & $\mathbf{4.61 \pm 0.07}$      
                       & $\mathbf{3.34 \pm 0.14}$ & $1.50 \pm 0.17$          & $2.84 \pm 0.12$              
                       & $3.72 \pm 0.09$          & $3.34 \pm 0.10$          & $\mathbf{3.88 \pm 0.09}$   \\
         Anger         & $3.86 \pm 0.09$          & $3.80 \pm 0.10$          & $\mathbf{3.93 \pm 0.08} $
                       & $\mathbf{4.31 \pm 0.09}$ & $3.88 \pm 0.06$          & $4.22 \pm 0.18 $ 
                       & $4.14 \pm 0.06$          & $3.97 \pm 0.08$          & $\mathbf{4.21 \pm 0.08} $    \\
         Amazement     & $3.58 \pm 0.10$          & $\mathbf{3.94 \pm 0.10}$ & $3.81 \pm 0.08$  
                       & $\mathbf{3.72 \pm 0.10}$ & $3.38 \pm 0.14$          & $3.50 \pm 0.15$    
                       & $3.83 \pm 0.07 $         & $3.62 \pm 0.08$          & $\mathbf{4.02 \pm 0.08}$      \\
         Disgust       & $3.42 \pm 0.08$          & $3.42 \pm 0.09$          & $\mathbf{3.64 \pm 0.08}$   
                       & $3.61 \pm 0.12$          & $3.88 \pm 0.06$          & $\mathbf{4.07 \pm 0.07}$   
                       & $3.97 \pm 0.09 $         & $3.70 \pm 0.09$         & $\mathbf{4.32 \pm 0.08}$   \\
         \midrule
         Average       & $3.43 \pm 0.04$          & $\mathbf{4.07 \pm 0.03}$          & $\mathbf{4.09 \pm 0.03}$   
                       & $\mathbf{3.76 \pm 0.04}$ & $3.04 \pm 0.08$          & $\mathbf{3.76 \pm 0.05}$    
                       & $4.03 \pm 0.03$          & $3.66 \pm 0.03$         & $\mathbf{4.10 \pm 0.03}$   \\
         \bottomrule
         
    \end{tabular}
    }
    \label{cmp_other_methods}
\end{table*}

\begin{table*}[htbp]
        \caption{Subjective evaluation results for three variants of iEmoTTS on cross-speaker emotion transfer. The Emotion similarity, speaker similarity and voice quality MOS with a 95\% confidence interval are compared. A higher MOS value indicates better performance. The MOS values significantly higher than other models are in bold.}
    \centering
    \scalebox{0.7}{
    \begin{tabular}{lcccc|cccc|cccc}
             \toprule
            \multirow{2}*{\textbf{Emotion}} & \multicolumn{4}{c}{Emotion Similarity MOS} & \multicolumn{4}{c}{Speaker Similarity MOS} & \multicolumn{4}{c}{Voice Quality MOS} \\
            \cmidrule(lr){2-5}\cmidrule(lr){6-9}\cmidrule(lr){10-13}
                       & iEmoTTS-NP               & iEmoTTS-SE                       & iEmoTTS-SENP   & iEmoTTS
                       & iEmoTTS-NP               & iEmoTTS-SE                       & iEmoTTS-SENP   & iEmoTTS 
                       & iEmoTTS-NP               & iEmoTTS-SE                       & iEmoTTS-SENP   & iEmoTTS \\
            \midrule

         Happiness     & $3.96 \pm 0.07$          & $3.84 \pm 0.07$                 & $3.89 \pm 0.07$       & $\mathbf{4.34 \pm 0.07}$
                       & $3.80 \pm 0.10$          & $\mathbf{3.94 \pm 0.09}$        & $3.83 \pm 0.08$       & $\mathbf{3.90 \pm 0.09}$
                       & $3.70 \pm 0.10$          & $\mathbf{4.34 \pm 0.06} $       & $4.04 \pm 0.09$       & $4.30 \pm 0.07$ \\

         Sadness       & $2.60 \pm 0.09$          & $2.11 \pm 0.08$                 & $2.04 \pm 0.07$       & $\mathbf{3.88 \pm 0.06}$
                       & $3.89 \pm 0.10$          & $\mathbf{3.94 \pm 0.10}$        & $3.74 \pm 0.10$       & $3.67 \pm 0.10$ 
                       & $3.73 \pm 0.09$          & $4.16 \pm 0.07$                 & $3.66 \pm 0.07$       & $\mathbf{4.24 \pm 0.07}$ \\
         Poorness      & $2.74 \pm 0.09$          & $2.46 \pm 0.09$                 & $2.45 \pm 0.08$       & $\mathbf{4.32 \pm 0.07}$
                       & $3.09 \pm 0.12$          & $\mathbf{3.97 \pm 0.10}$        & $3.75 \pm 0.07$       & $3.78 \pm 0.09$
                       & $3.20 \pm 0.10$          & $\mathbf{4.08 \pm 0.08}$        & $3.89 \pm 0.08$       & $3.73 \pm 0.06$      \\
         Fear          & $2.49 \pm 0.10$          & $2.35 \pm 0.09$                 & $2.32 \pm 0.07$       & $\mathbf{4.48 \pm 0.06}$
                       & $2.83 \pm 0.10$          & $\mathbf{3.88 \pm 0.09}$        & $3.70 \pm 0.07$       & $2.77 \pm 0.10$  
                       & $2.76 \pm 0.09$          & $\mathbf{4.04 \pm 0.07}$        & $3.46 \pm 0.07$       & $3.80 \pm 0.07$ \\
         Anger         & $3.78 \pm 0.08$          & $3.61 \pm 0.08$                 & $3.75\pm 0.08$        & $\mathbf{3.91\pm 0.07} $
                       & $3.98 \pm 0.10$          & $\mathbf{4.09 \pm 0.10}$                 & $3.86 \pm 0.09$       & $\mathbf{4.11 \pm 0.11} $ 
                       & $3.88 \pm 0.09$          & $\mathbf{4.25 \pm 0.07}$        & $3.85 \pm 0.07$       & $4.17 \pm 0.09 $  \\
         Amazement     & $3.26 \pm 0.10$          & $3.12 \pm 0.08$                 & $3.25 \pm 0.08$       & $\mathbf{3.84 \pm 0.07}$
                       & $3.77 \pm 0.11$          & $\mathbf{4.10 \pm 0.08}$        & $4.00 \pm 0.07$       & $3.40 \pm 0.11$    
                       & $3.12 \pm 0.10$          & $\mathbf{4.05 \pm 0.09}$        & $3.62 \pm 0.09$       & $4.00 \pm 0.06$  \\
         Disgust       & $\mathbf{3.51 \pm 0.08}$ & $3.49 \pm 0.08$                 & $3.48 \pm 0.08$       & $\mathbf{3.55 \pm 0.09}$
                       & $4.03 \pm 0.11$          & $\mathbf{4.09 \pm 0.09}$        & $3.91 \pm 0.08$       & $4.02 \pm 0.11$ 
                       & $3.54 \pm 0.09$          & $4.03 \pm 0.08 $                & $3.72 \pm 0.08 $      & $\mathbf{4.23 \pm 0.07}$   \\
         \midrule
         Average       & $3.19 \pm 0.04$          & $2.99 \pm 0.04 $                &  $3.03 \pm 0.03 $    & $\mathbf{4.05 \pm 0.03}$
                       & $3.63 \pm 0.04$          & $\mathbf{4.00 \pm 0.04}$        & $3.83 \pm 0.03$      & $3.66 \pm 0.05$
                       & $3.42 \pm 0.04$          & $\mathbf{4.14 \pm 0.03} $       & $3.75 \pm 0.08 $     & $4.07 \pm 0.03$\\
         \bottomrule
         
    \end{tabular}
    }
    \label{cmp_variants}
\end{table*}

\subsection{Effectiveness of Disentanglement}
Three variants of iEmoTTS, i.e., iEmoTTS-NP, iEmoTTS-SE and iEmoTTS-SENP, are compared to show the effectiveness of disentanglement between timbre and prosody. The three variants have similar structures to the standard iEmoTTS. However, they do not meet the conditions of prosody-timbre disentanglement described in \Cref{iemotts_model_overview}. More precisely, in iEmoTTS-NP, the prosodic features are not used in model input, and only emotion and phoneme encoding are combined before the up-sampling operation. Since the model is based on a duration-based TTS model, the prosody predictor is required, but only to predict duration information for the up-sampling layer. In iEmoTTS-SE, the timbre encoding is placed before the prosody predictor and combined with the phoneme and emotion encoding. The combined features are fed to the prosody predictor and then added with the prosodic information as the input to the speech decoder. In iEmoTTS-SENP, the timbre encoding is placed forward like in iEmoTTS-SE, while the prosodic features are removed like in iEmoTTS-NP. The other components and parameters of iEmoTTS-NP,  iEmoTTS-SE and iEmoTTS-SENP are the same as the iEmoTTS. 
In iEmoTTS-NP, the F0 and intensity information is absent from the model input, which might be absorbed by the timbre encoding. While in iEmoTTS-SE, the speaker and emotion factors are used together for prosody and mel-spectrograms predictions, which can be viewed as performing disentanglement between speaker and emotion. \autoref{cmp_variants} summarizes the evaluation results for the three variant models of iEmoTTS.

Regarding emotion similarity, the standard iEmoTTS achieves the highest scores on all emotion types compared with the three variant models. It is noted that the gaps between the MOS of iEmoTTS and the variant models are particularly significant in some emotion types, namely, sadness and poorness. This is possible because these emotion types have low arousal levels, while the utterances of target speakers generally have high arousal. In the training of iEmoTTS, the timbre encoding is forced to remove the prosodic information while keeping the timbre information. In the variant models, the timbre encoding would retain speaker-related prosodic information. Therefore, for cross-speaker emotion transfer, it is more challenging for the variant models to realize the emotion types absent in the speech of target speakers.

In terms of speaker similarity to the target speakers, iEmoTTS-SE shows significantly better performance than iEmoTTS, and iEmoTTS-SENP performs significantly better than iEmoTTS-NP. The results might be attributed to the fact that the perception of the speaker also involves prosody cues. In most cases, timbre plays a deterministic role in speaker perception. However, suppose the prosody of a synthesized utterance deviates largely from recordings of the target speakers. In that case, the listeners might consider the utterances to have lower speaker similarity to the target speakers. For example, when we transfer the emotion "fear" to a target speaker with the iEmoTTS model, the pitch level of the synthesized utterances is much lower than the mean pitch of that target speaker. Correspondingly, the speaker similarity MOS score of emotion ``fear'' is also much lower than other emotion types. Prior studies have also noted the importance of pitch level affecting listeners' perception of male and female voices \cite{kreiman2011foundations}. 

On voice quality,  iEmoTTS-SE demonstrates significantly better performance than iEmoTTS, and iEmoTTS-SENP performs significantly better than iEmoTTS-NP. This result could also be explained by the fact that the prosodic features predicted from iEmoTTS and iEmoTTS-NP deviated from the prosodic features in the recordings of target speakers. The unseen or deviated prosodic features might degrade the model performance on voice quality. The less satisfactory performance of iEmoTTS-NP compared with iEmoTTS can be attributed to the absence of F0 and intensity. This result is consistent with those reported in \cite{ren2020fastspeech, lancucki2021fastpitch}, where adding the F0 feature during training improves the quality of synthesized speech.

\subsection{Emotion Intensity Control}
\begin{figure*}[!t]
  \centering
  \includegraphics[width=0.9\textwidth]{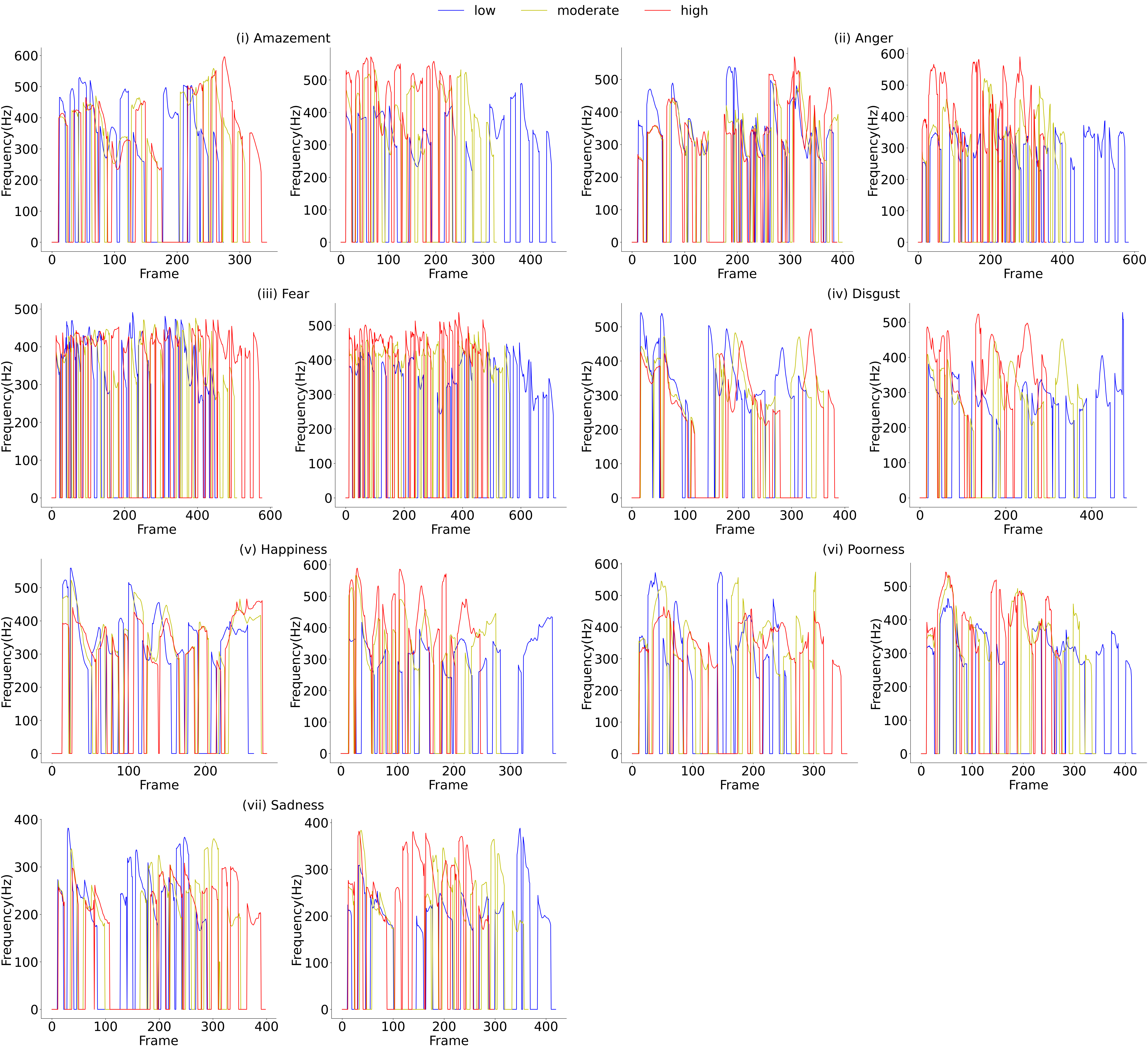}
  \caption{F0 curves of synthesized samples with high, moderate and low emotion intensity. The Left column represents the system iEmoTTS, while the right column represents the system iEmoTTS-S}
  \label{Figure:emotional_intensity_control}
\end{figure*}

In the proposed system, the emotion intensity of a given utterance is measured by the posterior probability that the utterance carries a specific emotion type. The intensity value is set to 0.1 for synthesizing speech with a `low' emotion intensity level and 1 for a `high' level. The median value of training data statistics is used to realize the `moderate' level. This method is compared with the method proposed as in \cite{li2021controllable,li2022cross}. In our experiment, the method described in \cite{li2021controllable,li2022cross} is implemented by removing the modified softmax layer in iEmoTTS. The resulting system is referred to as iEmoTTS-S. The iEmoTTS-S controls the emotion intensity by multiplying a scalar factor by the emotion encoding at the inference stage. The scalar is set to 0.5, 1.5, and 2.5 to represent the `low', `moderate', and `high' intensity levels. The emotion intensity ranking test results on iEmoTTS and iEmoTTS-S are shown as in \autoref{emotion_intensity}. To visualize the prosodic variation achieved by iEmoTTS and iEmoTTS-S, we plot the F0 curves of synthesized speech with different emotion intensity levels in \autoref{Figure:emotional_intensity_control}.

\begin{table*}[htbp]
        \caption{ Subjective evaluation results of the emotion intensity level ranking test for source and target speakers. The higher classification accuracy for each emotion intensity level indicates better performance. }
    \centering
    \scalebox{0.8}{
    \begin{tabular}{ccc|cc|cc|cc|cc|cc}
    \toprule
    \multirow{3}*{Emotion} & \multicolumn{6}{c}{Source Speakers}  & \multicolumn{6}{c}{Target Speakers} \\
    \cmidrule(lr){2-7}\cmidrule(lr){8-13}\cmidrule(lr){6-7}
    & \multicolumn{2}{c}{High} & \multicolumn{2}{c}{Moderate} & \multicolumn{2}{c}{Low} & \multicolumn{2}{c}{High} & \multicolumn{2}{c}{Moderate} & \multicolumn{2}{c}{Low} \\
    \cmidrule(lr){2-3}\cmidrule(lr){4-5}\cmidrule(lr){6-7}\cmidrule(lr){8-9}\cmidrule(lr){10-11}\cmidrule(lr){12-13}
               & iEmoTTS   & iEmoTTS-S   & iEmoTTS   & iEmoTTS-S & iEmoTTS   & iEmoTTS-S 
               & iEmoTTS   & iEmoTTS-S   & iEmoTTS   & iEmoTTS-S & iEmoTTS   & iEmoTTS-S \\
    \midrule
 Happiness     & $\mathbf{74\%}$ & $71\%$          &  $ 60\%          $  & $\mathbf{64\%}$ & $73\%$          & $\mathbf{83\%}$
               & $\mathbf{70\%}$ & $68\%$          &  $ 68\%          $  & $\mathbf{75\%}$ & $78\%$          & $\mathbf{88\%}$ \\
 Sadness       & $\mathbf{94\%}$ & $20\%$          &  $ \mathbf{90\%} $  & $52\%$          & $\mathbf{92\%}$ & $46\%$
               & $\mathbf{79\%}$ & $13\%$          &  $ \mathbf{93\%} $  & $36\%$          & $\mathbf{75\%}$ & $55\%$\\
 Poorness      & $\mathbf{98\%}$ & $46\%$          &  $ \mathbf{92\%} $  & $58\%$          & $\mathbf{92\%}$ & $54\%$ 
               & $\mathbf{88\%}$ & $21\%$          &  $ \mathbf{93\%} $  & $32\%$          & $\mathbf{88\%}$ & $65\%$ \\
 Fear          & $\mathbf{78\%}$ & $36\%$          &  $ \mathbf{76\%} $  & $49\%$          & $\mathbf{93\%}$ & $65\%$ 
               & $\mathbf{76\%}$ & $34\%$          &  $ \mathbf{84\%} $  & $41\%$          & $\mathbf{80\%}$ & $72\%$  \\
 Anger         & $\mathbf{86\%}$ & $85\%$          &  $ \mathbf{78\%} $  & $57\%$          & $\mathbf{84\%}$ & $80\%$ 
               & $71\%$          & $\mathbf{79\%}$ &  $ 60\% $  & $\mathbf{70\%}$          & $71\%$          & $\mathbf{88\%}$  \\
 Amazement     & $\mathbf{74\%}$ & $64\%$          &  $ \mathbf{76\%} $  & $51\%$          & $\mathbf{88\%}$ & $65\%$
               & $66\%         $ & $66\%$          &  $ \mathbf{68\%} $  & $51\%$          & $\mathbf{82\%}$ & $80\%$ \\
 Disgust       & $\mathbf{60\%}$ & $49\%$          &  $ 51\%          $  & $51\%$          & $70\%$          & $\mathbf{71\%}$
               & $52\%$          & $\mathbf{54\%}$ &  $ \mathbf{57\%} $  & $54\%$          & $51\%$          & $\mathbf{78\%}$ \\
     \midrule
     Average   &$\mathbf{81\%}$  & $53\%$          &  $ \mathbf{75\%} $  & $55\%$          & $\mathbf{85\%}$          & $67\%$
               &$\mathbf{72\%}$  & $48\%$          &  $ \mathbf{74\%} $  & $51\%$          & $75\%$          & $75\%$\\
    \bottomrule
    \end{tabular}
    
    }
    \label{emotion_intensity}
\end{table*}

The evaluation results reveal that the proposed method is able to successfully and consistently realize different levels of emotion intensity. A closer inspection on \autoref{emotion_intensity} shows that the performance is better on the utterances of source speakers than those of target speakers. A possible explanation is that other speech factors apart from prosody may also contribute to the perceived emotion intensity. The speech of source speakers has a broader range of emotion intensity than that of target speakers. With iEmoTTS, prosodic features can be obtained with any value of emotion intensity for speech generation with the target speakers. However, there is no appropriate manipulation of other speech factors at the same time \cite{schuller2013computational}.

The proposed method is better than the scalar-based emotion intensity control method in three aspects. First, our method introduces subtle and diverse variations for different emotion types. Specifically, higher emotion intensity leads to a slower speaking rate for the emotion types of `sadness', `fear', `poorness', and `disgust'.  For `amazement', higher intensity results in a low F0 at the beginning of a sentence and a high F0 and slow speaking rate at the end to create a feeling of surprise. Higher emotion intensity applied to `fear" produces speech with an abrupt lifting of F0 at the end of the first phrase. For the scalar-based emotion intensity control method used in iEmoTTS-S, synthesized speech with low emotion intensity for all emotion types comes with low pitch and slow speaking rate. High emotion intensity brings high pitch values and a fast speaking rate for all emotion types. 

Second, in most cases, the proposed method achieves better classification performance in assigning (ranking) the utterances to correct intensity levels. The superiority is particularly evident in `fear', `sadness', and `poorness'. One unexpected finding is that the scalar-based method could perform nearly on par with our method on some emotion types with high arousal levels, e.g., happiness and anger. The scalar-based method generates speech with similar prosodic variations at the same emotion intensity level. The comments collected from the test participants indicate that speech with relatively low pitch and slow tempo are often considered as low intensity, and high pitch and fast tempo as high intensity for emotion types with high arousal, even if the test utterances do not sufficiently express the desired emotion.

Third, the voice quality of speech generated by iEmoTTS is better than iEmoTTS-S. In the training of iEmoTTS-S, the emotion intensity value is always set to 1. Whilst at the inference stage, iEmoTTS-S uses other values of intensity. The degraded quality could be attributed to the mismatch between training and inference.

\subsection{Zero-shot Cross-speaker Emotion Transfer}
The performance of iEmoTTS on zero-shot cross-speaker emotion transfer is evaluated in this section. In addition to the default setting, we evaluate different information bottleneck (IB) capacities on the bottleneck layer in the timbre encoder. More specifically, IB-full refers to the bottleneck layer removed after the speaker encoder. All information in the speaker vector is fed directly to the model with no constraint. IB-large is the system obtained by setting the group number in the modified VQ-VAE to 8, which is larger than the default setting. Therefore, the IB capacity of the bottleneck layer is larger, allowing more information to be passed through the bottleneck layer. IB-small works with a smaller IB capacity with a group number of 2. Subjective evaluation on emotion similarity and speaker similarity are detailed in \autoref{mos_zero}.

\begin{table*}[htbp]
        \caption{The emotion similarity and speaker quality MOS of four settings for iEmoTTS with zero-shot cross-speaker emotion transfer with 95\% confidence interval. The MOS values significantly higher than other models are in bold.} 
    \centering
    \scalebox{0.9}{
    \begin{tabular}{lcccc|cccc}
             \toprule
            \multirow{2}*{Emotion} & \multicolumn{4}{c}{Emotion Similarity MOS}  & \multicolumn{4}{c}{Speaker Similarity MOS} \\
            \cmidrule(lr){2-5}\cmidrule(lr){6-9}
                       &  IB-full                 &  IB-large                & IB-default             & IB-small            
                       &  IB-full                 &  IB-large                & IB-default             & IB-small  \\
            \midrule

         Happiness     & $4.00 \pm 0.09$          & $\mathbf{4.43 \pm 0.12}$ & $4.37 \pm 0.08$         & $4.31 \pm 0.08$   
                       & $3.27 \pm 0.15$          & $\mathbf{3.37 \pm 0.11}$          & $3.34 \pm 0.10$         & $\mathbf{3.35 \pm 0.11}$ \\ 
         Sadness       & $4.29 \pm 0.09$          & $3.96 \pm 0.12$          & $\mathbf{4.37 \pm 0.08}$         & $\mathbf{4.38 \pm 0.08}$  
                       & $\mathbf{3.51 \pm 0.15}$          & $3.44 \pm 0.10$          & $\mathbf{3.52 \pm 0.10}$& $3.11 \pm 0.11$   \\
         Poorness      & $3.20 \pm 0.13$          & $3.02 \pm 0.11$          & $3.33 \pm 0.11$         & $\mathbf{3.47 \pm 0.11}$    
                       & $\mathbf{3.11 \pm 0.13}$ & $2.98 \pm 0.11$          & $3.05 \pm 0.11$         & $2.74 \pm 0.11$    \\
         Fear          & $4.03 \pm 0.10$          & $3.66 \pm 0.13$          & $\mathbf{4.08 \pm 0.11}$& $\mathbf{4.07 \pm 0.12}$           
                       & $\mathbf{2.78 \pm 0.13}$ & $2.35 \pm 0.09$          & $2.52 \pm 0.09$         & $2.20 \pm 0.09$ \\ 
         Anger         & $4.35 \pm 0.07$          & $\mathbf{4.68 \pm 0.09}$ & $4.49 \pm 0.07$         & $\mathbf{4.65 \pm 0.07}$    
                       & $3.50 \pm 0.14$          & $\mathbf{3.56 \pm 0.10}$ & $3.52 \pm 0.10$         & $3.51 \pm 0.11$     \\
         Amazement     & $3.62 \pm 0.10$          & $\mathbf{3.95 \pm 0.12}$ & $3.91 \pm 0.10$         & $3.85 \pm 0.10$   
                       & $\mathbf{3.41 \pm 0.15}$ & $3.37 \pm 0.15$          & $3.20 \pm 0.10$         & $3.05 \pm 0.11$    \\
         Disgust       & $3.62 \pm 0.10$          & $3.62 \pm 0.16$          & $3.94 \pm 0.10$         & $\mathbf{3.97 \pm 0.11}$   
                       & $3.37 \pm 0.14$          & $\mathbf{3.51 \pm 0.15}$ & $3.11 \pm 0.11$         & $2.85 \pm 0.12$   \\
         \midrule
         Average       & $3.87 \pm 0.04$          & $3.90 \pm 0.05$          & $4.07 \pm 0.03$         &  $\mathbf{4.10 \pm 0.04}$    
                       & $\mathbf{3.28 \pm 0.05}$ & $3.23 \pm 0.06$          & $3.18 \pm 0.04$         &  $2.97 \pm 0.04$    \\
         \bottomrule
         
    \end{tabular}
    }
    \label{mos_zero}
\end{table*}

The results suggest that synthesized speech with the default IB setting achieves significantly better performance on emotion similarity than other settings and maintains good speaker similarity on the zero-shot cross-speaker emotion transfer task. It is interesting to note that the MOS on emotion similarity increases as the IB capacity decreases. The growth of emotion similarity is attributed to the narrowing bottleneck, which allows only the information pertinent to the timbre to remain in the timbre encoding. Compared with the case of seen target speakers (\autoref{cmp_other_methods}), IB-default and IB-small settings can achieve comparable performance on unseen target speakers in the zero-short scenario. As the IB capacity decreases, the MOS on speaker similarity decreases. This may be explained by the model removing some of the timbre information when the information bottleneck is made narrow. It is also noted that the MOS on speaker similarity decreases slightly from IB-full to IB-default and drops abruptly from IB-default to IB-small setting. As the IB capacity shrinks, the model would be biased to discard most other information and a small portion of timbre information. If the information bottleneck is too narrow, the model may discard too much timbre information. This suggests a trade-off between emotion and speaker similarity as the IB capacity varies. 

\subsection{Effectiveness of Emotion Encoder}
The benefits of the proposed emotion encoder are evaluated in this section. Two variants of the iEmoTTS are created for comparison, namely iEmoTTS-SER and iEmoTTS-WoEI. In iEmoTTS-SER, a speech emotion recognition (SER) model was first trained with utterances of the source speakers. The SER model follows the structure described in \autoref{analysis_emotion}. Each utterance from target speakers is predicted with an emotion label by inputting its mel-spectrograms into the SER model. Since all utterances have either ground-truth or predicted emotion labels, the emotion encoder and emotion intensity module are removed from iEmoTTS. In iEmoTTS-WoEI, the emotion intensity module in the emotion encoder of the iEmo-TTS is removed such that the output of the emotion encoder contains only the discrete emotion type ID. The other components and parameters of iEmoTTS-SER and iEmoTTS-WoEI remain the same as iEmoTTS. The evaluation results on emotion similarity and voice quality are shown as in \autoref{effectiveness_emo_enc}.

\begin{table*}[htbp]
        \caption{The emotion similarity and voice quality MOS for iEmoTTS-WoEI, iEmoTTS-SER and iEmoTTS with 95\% confidence interval. The MOS values significantly higher than other models are in bold.}
    \centering
        \scalebox{0.9}{
    \begin{tabular}{lccc|ccc}
             \toprule
            \multirow{2}*{Emotion type} & \multicolumn{3}{c}{Emotion Similarity MOS}  & \multicolumn{3}{c}{Voice Quality MOS} \\
            \cmidrule(lr){2-4}\cmidrule(lr){5-7}
                       & iEmoTTS-WoEI              & iEmoTTS-SER              & iEmoTTS          
                       & iEmoTTS-WoEI              & iEmoTTS-SER              & iEmoTTS  \\
            \midrule

         Happiness     & $4.30 \pm 0.08$          & $4.13 \pm 0.08$          & $\mathbf{4.33 \pm 0.07}$  
                       & $4.19 \pm 0.06$          & $\mathbf{4.25 \pm 0.07}$ & $4.23 \pm 0.08$ \\
         Sadness       & $2.95 \pm 0.10$          & $2.85 \pm 0.10$          & $\mathbf{3.85 \pm 0.06}$ 
                       & $4.26 \pm 0.06$          & $\mathbf{4.30 \pm 0.07}$ & $4.18 \pm 0.08$  \\
         Poorness      & $3.97 \pm 0.10$          & $4.03 \pm 0.08$          & $\mathbf{4.19 \pm 0.08}$
                       & $\mathbf{3.59 \pm 0.09}$ & $3.52 \pm 0.09$          & $\mathbf{3.59 \pm 0.09}$              \\
         Fear          & $4.08 \pm 0.09$          & $4.30 \pm 0.07$          & $\mathbf{4.44 \pm 0.07}$      
                       & $3.82 \pm 0.08$          & $3.83 \pm 0.09$          & $\mathbf{3.87 \pm 0.09}$   \\
         Anger         & $\mathbf{4.09 \pm 0.09}$ & $3.69 \pm 0.08 $         & $3.81 \pm 0.06 $
                       & $4.25 \pm 0.05$          & $\mathbf{4.29 \pm 0.08}$ & $4.15 \pm 0.08 $    \\
         Amazement     & $3.27 \pm 0.11$          & $3.17 \pm 0.09$          & $\mathbf{3.63 \pm 0.08}$  
                       & $\mathbf{4.07 \pm 0.06}$ & $4.01 \pm 0.08$          & $3.96 \pm 0.08$      \\
         Disgust       & $\mathbf{3.71 \pm 0.09}$ & $3.51 \pm 0.08$          & $3.51 \pm 0.07$   
                       & $4.14 \pm 0.06$          & $4.02 \pm 0.08$          & $\mathbf{4.20 \pm 0.08}$   \\
         \midrule
         Average       & $3.77 \pm 0.04$          & $3.67 \pm 0.04$          & $\mathbf{3.97 \pm 0.03}$   
                       & $\mathbf{4.05 \pm 0.03}$ & $4.03 \pm 0.03$          & $4.02 \pm 0.03$   \\
         \bottomrule
         
    \end{tabular}
    }
    \label{effectiveness_emo_enc}
\end{table*}

The three systems show similar performance on voice quality. iEmoTTS-WoEI is better than iEmoTTS-SER in the aspect of emotion similarity. The emotion encoder in iEmoTTS-WoEI is jointly trained with other components. It is believed that such end-to-end training can better model emotion-related information in mel-spectrograms. The results also show that iEmoTTS achieves significantly higher emotion similarity than iEmoTTS-WoEI. The gain could come from better emotion representation in iEmoTTS, where both emotion type ID and emotion intensity are utilized in model training.

\subsection{Effectiveness of Semi-supervised Strategy}
In this section, we evaluate the benefits of the semi-supervised strategy. In iEmoTTS, the emotion types for the utterances of target speakers are derived from semi-supervised learning and do not necessarily to be neutral. Two variants of iEmoTTS are created for comparison, namely iEmoTTS-NEU and iEmoTTS-NEU2, where the semi-supervised strategy is not applied, and utterances of target speakers are provided with neutral labels. The iEmoTTS-NEU and iEmoTTS-NEU2 have the same model structure as iEmoTTS. The only difference between the two variants is the neutral labels for target speakers.
In iEmoTTS-NEU, the neutral label for target speakers is considered identical to that for source speakers. In iEmoTTS-NEU2, the target speakers are given a new neutral label, which is to be differentiated from the neutral label for source speakers. This variant is motivated by \cite{li2022cross}, where the model treats the labels of target speakers as a new type of neutral style. The evaluation results on emotion similarity and voice quality are shown in \autoref{effectiveness_semi}.

\begin{table*}[htbp]
        \caption{The emotion similarity and voice quality MOS for iEmoTTS-NEU, iEmoTTS-NEU2, and iEmoTTS with 95\% confidence interval. The MOS values significantly higher than other models are in bold.}
    \centering
           \scalebox{0.9}{
    \begin{tabular}{lccc|ccc}
             \toprule
            \multirow{2}*{Emotion type} & \multicolumn{3}{c}{Emotion Similarity MOS}  & \multicolumn{3}{c}{Voice Quality MOS} \\
            \cmidrule(lr){2-4}\cmidrule(lr){5-7}
                       & iEmoTTS-NEU             & iEmoTTS-NEU2              & iEmoTTS           
                       & iEmoTTS-NEU             & iEmoTTS-NEU2              & iEmoTTS  \\
            \midrule

         Happiness     & $\mathbf{4.51 \pm 0.06}$ & $4.49 \pm 0.06$          & $\mathbf{4.51 \pm 0.06}$  
                       & $4.07 \pm 0.07$          & $\mathbf{4.25 \pm 0.06}$ & $\mathbf{4.25 \pm 0.08} $ \\
         Sadness       & $\mathbf{4.49 \pm 0.08}$ & $4.33 \pm 0.07$          & $4.27 \pm 0.09$ 
                       & $3.66 \pm 0.08$          & $3.82 \pm 0.08$          & $\mathbf{4.12 \pm 0.08}$  \\
         Poorness      & $4.13 \pm 0.10$          & $\mathbf{4.22 \pm 0.10}$ & $4.18 \pm 0.06$
                       & $3.44 \pm 0.08$          & $3.32 \pm 0.08$          & $\mathbf{3.54 \pm 0.09}$              \\
         Fear          & $4.42 \pm 0.08$          & $4.39 \pm 0.09 $         & $\mathbf{4.53 \pm 0.07}$      
                       & $3.93 \pm 0.09 $         & $3.92 \pm 0.09$          & $\mathbf{3.97 \pm 0.09}$   \\
         Anger         & $\mathbf{3.99 \pm 0.07}$ &$3.94 \pm 0.08 $          & $3.93 \pm 0.07 $
                       & $4.08 \pm 0.07 $         & $4.12 \pm 0.07$          & $\mathbf{4.19 \pm 0.08} $    \\
         Amazement     & $3.92 \pm 0.09 $         & $3.81 \pm 0.10 $         & $\mathbf{3.99 \pm 0.07}$  
                       & $3.95 \pm 0.07 $         & $3.82 \pm 0.07$          & $\mathbf{4.10 \pm 0.08}$      \\
         Disgust       & $3.64 \pm 0.10 $         & $3.67 \pm 0.10 $         & $\mathbf{3.71 \pm 0.08}$   
                       & $\mathbf{4.31 \pm 0.06}$ & $4.25 \pm 0.07$          & $4.30 \pm 0.08$   \\
         \midrule
         Average       & $\mathbf{4.16 \pm 0.03}$ & $4.12 \pm 0.04$          & $\mathbf{4.16 \pm 0.03}$   
                       & $3.92 \pm 0.03 $         & $3.93 \pm 0.03$          & $\mathbf{4.07 \pm 0.03}$   \\
         \bottomrule
         
    \end{tabular}
    }
    \label{effectiveness_semi}
\end{table*}

There is no significant difference between the three systems in the aspect of emotion similarity. However, iEmoTTS achieve better voice quality than both iEmoTTS-NEU and iEmoTTS-NEU2. This reveals the effectiveness of the semi-supervised training strategy. In the training of iEmoTTS-NEU and iEmoTTS-NEU2, emotion labels and target speaker ID combinations are not seen. The generated mel-spectrograms tend to be degraded in the synthesis stage.

\section{Conclusions}
An end-to-end TTS model for a cross-speaker emotion transfer system has been developed based on timbre-prosody disentanglement. The system, iEmoTTS, encodes emotion-related information in speech in terms of a discrete emotion type and a probability-based emotion intensity value. It is able to generate emotional speech for seen or unseen target speakers via a process of cross-speaker emotion transfer. Notably, the cross-speaker emotion transfer can be done even if the speaker and emotion are highly entangled in the given speech data of source speakers. Extensive experiments on subjective evaluation have been performed to assess iEmoTTS in various aspects. The evaluation results show that iEmoTTS performs better than a variety of reference systems and variants of itself. iEmoTTS shows positive results on zero-shot transfer to unseen target speakers.With a properly chosen information bottleneck capacity, the proposed model is able to achieve a good balance between speaker similarity and emotion similarity. Lastly, the proposed method of emotion intensity control has been shown effective such that the desired intensity level of speech emotion can be well recognized by human listeners.

\ifCLASSOPTIONcaptionsoff
  \newpage
\fi




\bibliographystyle{IEEEtran}
\bibliography{IEEEabrv,Bibliography}

\vfill


\end{document}